\documentclass[letterpaper,12pt]{article}

\usepackage{hyperref}
\usepackage{color,soul}
\usepackage{amsmath}
\usepackage{rotating}
\usepackage{multicol}

\usepackage{mathptmx}

\usepackage{graphics}

\usepackage[authoryear,comma,longnamesfirst,sectionbib]{natbib}

\makeatletter
\newcommand{\@supervisor}{}
\newcommand{\supervisor}[1]{\renewcommand{\@supervisor}{#1}}
\newcommand{\@institute}{}
\newcommand{\institute}[1]{\renewcommand{\@institute}{#1}}
\makeatother

\title{Causal Hangover Effects}
\author{Andreas Santucci, Eric Lax}
\institute{Stanford University}
\supervisor{Guido Imbens}

\makeatletter
\renewcommand*{\maketitle}{%
    \begin{titlepage}
        \begin{center}

        \huge\@title\\
        \vspace{1.5cm}
        \Large\@author
        \vspace{1.5cm}

        \vfill

        \normalsize
        Supervised by:\\
        \@supervisor

        \vspace{0.8cm}

        \normalsize
        \@institute

        \end{center}%
    \end{titlepage}%
}
\makeatother

\begin{document}
\maketitle


\section{Introduction}
It's not unreasonable to think that in-game sporting performance can be affected 
partly by what takes place off the court. 
We can't observe what happens between games directly. Instead, we proxy for the 
possibility of athletes partying by looking at play following games in party cities.
We are interested to see if 
teams exhibit a decline in performance the day following a game in a city 
with active nightlife; we call this a ``hangover effect''.
Part of the question is determining a reasonable way to measure
levels of nightlife, and correspondingly which cities are notorious for it; we colloquially refer to such cities as ``party cities''.
At first glance, the topic might seem callow, but in fact there are many interesting
nuances to work through.

Inquiring about this causal relationship is worthwhile for several reasons. 
Knowing how player performance is affected by external factors could inform coaches of a more
optimal play strategy: if visitation to a city with active nightlife is correlated with
next-day injury, perhaps coaches might bench their star players if the latter game isn't crucial.
It could also inform team managers how travel schedules should be set, or even team rules.
Further, there are potential opportunities for arbitrage.  The
NBA Commissioner Adam Silver has reported that there is \$400 billion per year spent 
on sports betting in the U.S. alone.\citep{asilver,cnnsilver} 

How do we carry out this study?
We exploit data on bookmaker spreads: the expected score 
differential between two teams after conditioning on observable performance in past games
and expectations about the upcoming game. 
We expect a team to meet the spread half the time, since this is one of the easiest ways for 
bookmakers to guarantee a profit.\citep{dubnerlevitt} We construct a model which attempts to 
estimate the causal effect of  visiting a ``party city'' on subsequent day 
performance as measured by the odds of beating the spread. In particular, we only consider the hangover effect 
on games played back-to-back within 24 hours of each other.
To the extent that odds of beating the spread against next day opponent 
is uncorrelated with playing in a party city the day before, which should be the case 
under an efficient betting market, we have identification in our variable of interest.

We perform this analysis for both the National Basketball Association
(NBA) and Major League Baseball (MLB). 
We find that visiting a city with active nightlife the day prior to a game does have a 
statistically significant negative effect on a team's likelihood of meeting bookmakers' expectations
for both NBA and MLB. Within the NBA, we analyze specific team performance metrics such as number of points allowed, and find that it is 
negatively effected by previous day visitation to a city with active nightlife. Having established a causal model, we realize a profitable betting scheme for MLB. We also perform a robustness check and see that if players rest more than
 24 hours after visiting a ``party city'', the hangover effect dissipates.

\section{Effects of Partying on Athletic Performance}

\subsection{Anecdotal evidence of the effects on sporting performance}
Professional players openly admit they are interested in nightlife: Matt Barnes recently complained that 
``there's no nightlife in Utah'' when asked about playing the Jazz in the second round of NBA playoffs 
in 2017.\citep{cestone}
There are also numerous reports of teams partying to the detriment of the game.
Within the NBA, J.R. Smith's performance metrics substantially improved 
 after being traded from New York to Cleveland; when an interviewer
asked what contributed to this bump in performance, he infamously replied
that in New York he would have been out partying, but because there's no nightlife in Cleveland he ends up hitting the gym.\citep{price,ley} Before being traded, J.R. Smith's career-to-date success rate of shots taken was 42 percent, but somehow he only managed 
34 percent on Sunday afternoon games in the 2012-13 season.
Presumably, this was because such games followed a long weekend of partying.\citep{princeofthecity} It has been suggested that because
NBA games are played in the evenings, athletes can get away with late-nights.\citep{drodmanbleacherreport}

\subsection{Individual players}
Studies have also been carried out on statistical ranking of the players
most notorious for pursuing active nightlife in the NBA.\citep{chase} In particular, 
the author rejects the null hypothesis that J.R. Smith shoots just as well on Sundays as he does
every other day of the week at the $p=0.01$ level of significance; however, the author also
cautions that with so many players in the NBA, it is expected that we find a split as extreme
as J.R.'s simply due to chance.
The study goes on to look
at historical performance for all NBA players who began their career after 1974, and compares
Sunday afternoon performance to overall career performance. 
In particular, the study examines
2,035 players with at least one shot on Sunday games and finds that 1,092 of them shot worse
on Sundays than they did on all other days, as measured by the ratio of shots scored to shots
taken. A binomial test concludes that across the NBA, 
individual players perform worse on Sundays. Effect sizes are not considered, 
i.e. although players perform worse, we don't know by how much.

\subsection{Teams}
There has been an attempt to answer the question at the team-level before: i.e. do teams perform
worse after a night on the town, and if so does it depend on which city they 
visited.\citep{ezekowitz} However, the study only
looks at performance in party cities, as opposed to subsequent day performance (i.e. the study does not consider a delayed hangover effect). Moreover, the sample size was much smaller than ours, because the author only considered Sunday afternoon games from 2006-14. Finally, there was no methodology used behind the selection of party cities: the authors first define ad-hoc five ``party cities'', and when that did not yield a significant result they estimate fixed-effects for each city. The author does carry out a binomial test, using 284 games, of how many home-game covers vs. away-game covers were met; the author concludes with weak statistical evidence that teams  systematically play worse on Sunday afternoons when on the road, 
regardless of city, as measured by their ability to meet bookmakers' expectations.

Allow us to elaborate on why we think that a delayed hangover effect is a more plausible model.
Suppose for a moment that teams do in fact systematically play worse when playing with cities in active nightlife.
Notice that a simple fixed effect for the team match-up and an interaction with a home team indicator would be sufficient to model this.
Given the size of the betting market, it's not hard to believe that such a simple arbitrage opportunity has already been considered
and factored into bookmakers' expectations. There is also the issue of an unknown travel schedule: if a team plays in a party city,
we can't infer whether they even had the option of being exposed to nightlife without looking at where they were the day prior.
This motivates us to examine a feature describing lagged game location.

\subsection{Laboratory studies}
Studies on effects of alcohol on sports performance in a controlled setting
show that alcohol impairs the nervous system, resulting in a decline in both
cognitive function and motor skill.\citep{shirreffs} In a study on 
the effects of alcohol on recovery of male athletes, it was found that
consumption is correlated with reduced production of testosterone
which ultimately leads to inhibited ability to recover and adapt to exercise.\citep{mjbarnes}
Sleep deprivation has also been shown to strongly impair cognitive and motor performance.\citep{pilcher} Both alcohol and sleep deprivation are well correlated with nightlife.

\section{Defining a Measure for Partying}

\paragraph{On the correlation between next day opponent and last game location}
A key assumption in our analysis is that next-day opponent is reasonably uncorrelated with
last game location. We plot a heat-map displaying the relationship between last game location
and next game location, and further break this down according to whether the game(s) were played at home or away.

\begin{figure}
  \centering
  \label{fig: next day opponent}
  \includegraphics[scale=0.45]{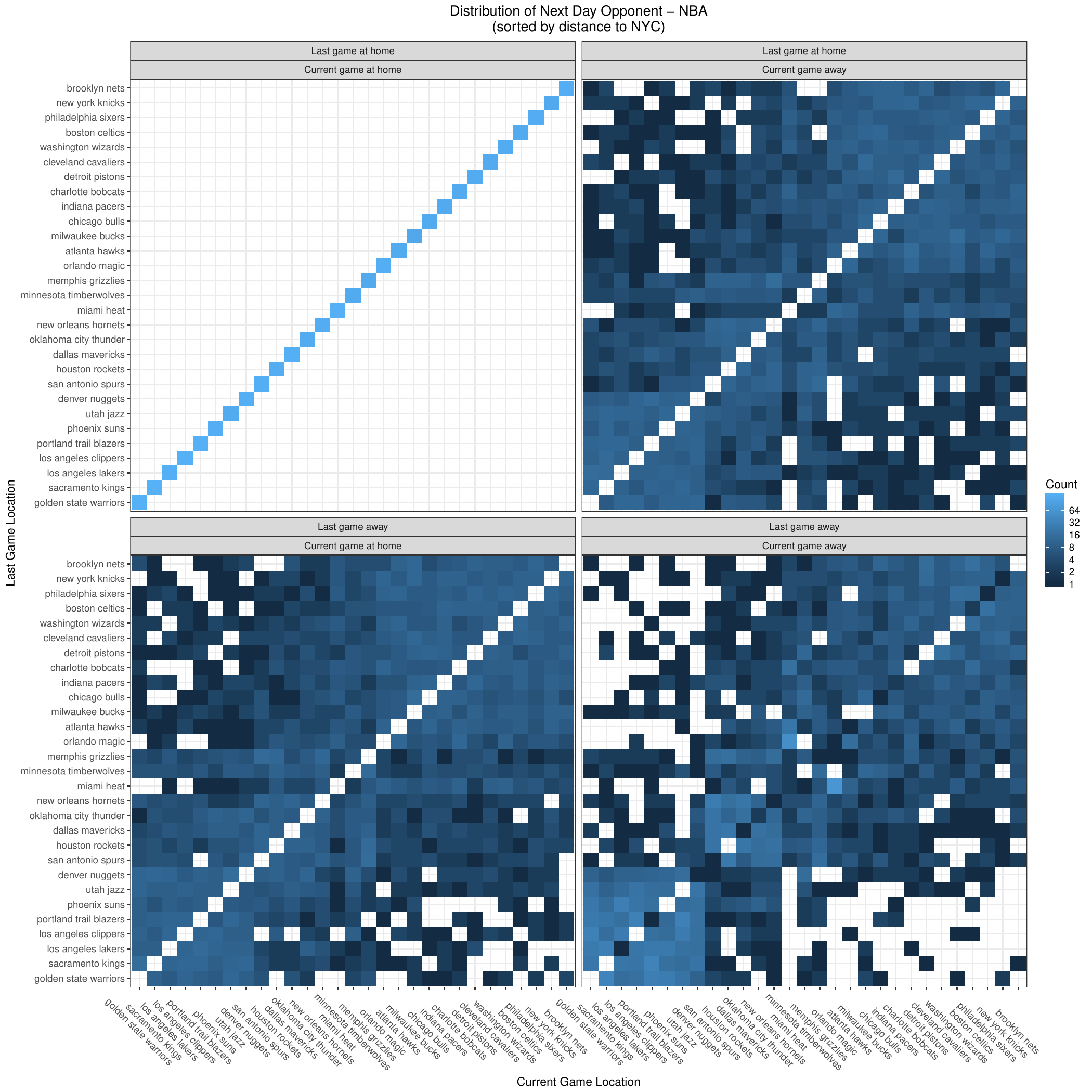}
  \caption{We examine the relationship between last game location and next day opponent in the NBA, sorted by distance from New York City. 
Top left: notice that the most likely scenario across all seasons is that a game will be played at home, and the last game was also at home. This makes sense; 1/2 of all games are played at home, and 1/2 are played away, so clearly each team is most likely to be found playing a home game for any given game. When exactly one of the last game or current game was played away,
we observe that some combinations cease to exist in our data; e.g. we never observe Charlotte Bobcats play Utah Jazz on the road and then fly directly back home for a game. Lastly, notice the patterns that emerge when both the last game and current game are played on the road. E.g. we see that when last game location was in Los Angeles, next day opponent likely to be a west coast team. There appears to be less of a correlation between last game location in New York or Brooklyn and next day opponent, perhaps because flying from east-to-west
is less burdensome as it relates to timezone changes.} 
\end{figure}

The figure displays the a count of co-occurences between last game location and current game location, faceted by home and away games.
Our takeaway from this figure is that indeed there is some correlation between last game location
and next day opponent (more so when both are away games,  
and in particular this is true for west coast teams). We hypothesize that
this is due to practical limitations on flight schedules. E.g. if a team is unwilling to take a red-eye
flight, then there is little hope to play an evening west coast game and make it to the east coast for a game the very next day;
the same is \emph{not true} when flying from east to west, in which case a morning flight is reasonable.

The fact that there is some correlation between last game location and next day opponent, even if only in west coast cities, increases the motivation for conditioning on spread, which may help account for systematic biases to the extent that some of these next-day opponents are particularly (not) skilled. 
In fact, using point-spread data we may weaken our assumption: we only need
likelihood of meeting the spread to be uncorrelated with last game location
in order to have identification in our variable of interest.

\paragraph{Partying as a latent variable}
One of the hardest parts of this analysis is that partying is a latent variable.
In order to determine when players partied, one may consider looking at social media data for direct mentions of partying, either by athletes or witnesses. However, due to the intense public scrutiny that athletes face, there is a strong reluctance to disclose their personal lives on social media. There are notable exceptions, such
as Oddel Beckham Jr.'s trip to Miami in the 2016-17 season.\citep{bieler, bleler}
However, these reports are few and far enough between that picking up statistical evidence would be fruitless.
Instead, we develop two approaches which do not rely on this data: a discrete indicator 
based on (a recently retired) player interview and also a continuous measure of nightlife based on local community patterns.

\paragraph{A discrete indicator}
Because famous athletes often enjoy a celebrity lifestyle, using conventional statistics
such as number of drinking establishments in each city or population does not work (see appendix \ref{fig: why drinks or pop dont work}), athletes instead
tend to visit high end nightclubs. We again turned to player interviews for ideas,
and found two major insights. From our conversations, 
we obtained a listing of cities which were anecdotally
mentioned as having active night-life. Between players, the rank ordering of these
was consistent to the extent that Los Angeles and New York city were top. 
This motivates using a binary indicator for party when either of these two cities have been visited within 24-hours of another game. 

\paragraph{A continuous measure}
We also found that athletes tend to party with other A-list celebrities,
particularly musicians.\citep{jbutlerdwade} After winning the NBA finals
in 2016, the Cleveland Cavaliers were found celebrating in nightclubs.\citep{xsclub} 
This motivated a continuous measure of nightlife, based on a simple count of the number of musicians in the metropolitan statistical area (MSA) 
in which the stadium resides: if the last game was played at most 24-hours ago, use the log of the total number
of musical groups, sound recording studios, and musical publishers in the MSA of the 
last game location, else emit zero.  We also re-scale this feature to lie within the
interval $[0,1]$ such that the magnitude of the coefficient is comparable with our
discrete indicator of nightlife.
This captures how much nightlife there is in a city since musical performances are correlated
with parties, if not an essential feature.

\section{Data}
\subsection{Points-spread and money-lines data} We first look toward 
\href{http://www.basketball-reference.com/leagues/}{basketball-reference.com} and
\href{http://www.baseball-reference.com/leagues/}{baseball-reference.com}
for a listing of game-days by season. With these listings, we queried \href{http://www.covers.com/sports/NBA/matchups?selectedDate=2011-1-01}{covers.com},
for all games occurring on a particular day, and for each we observe the
outcome of each game alongside the either the point-spread (for NBA) or money-line (for MLB) 
set by a betting house. A point-spread describes the expected score differential
between two teams, and a money-line describes the odds that a
team will win a match-up.

We interpret the point-spread according to a standard guide.\citep{hrisports} 
The reason we utilize bookmaker spreads is because this measure already incorporates
up-to-date information on how well the team has been performing in said season and also
their expected (dis)-advantage relative to a particular opponent. Without a feature
describing bookmaker expectations, we would need to construct a separate model (from scratch)
which estimates the probability that one team will be victorious over another, for all
$\binom{n}{2}$ possible match-ups.

\paragraph{NBA}
We obtain
data for seasons 2010-11 through 2016-17; in basketball, the season runs from October through April. 
There are exactly 30 teams in the NBA,
each plays 82 games per season.
We are missing the first couple months of games 
in the 2010-11 season, and our 2011-12 season has 229 missing games. In each remaining
season, we have at most two games missing of the $82 \times 30 / 2 = 1,230$ played.
We have a total of 7,829 games in our data-set. We add features
to our data such as number of days since last game, lagged game location, and 
travel distance.

\paragraph{MLB} We also obtain data for seasons 2011 through 2017; each season runs from 
April through October. There are also exactly 30 teams in the MLB, each plays 162 games per season.
We don't have all $162 \times 30 / 2 = 2,430$ games played each season, but instead only have
on average almost 2,100 games per season. We posit that the missing observations
are missing in a way that is exogenous to our model; a hand-inspection of several
reveals that the website URL was defunct.
We have a total of 12,709 games in our data-set. We again add features describing rest time and travel distance.


\subsubsection{Logic for using betting metrics}
In order to assert a causal relation between cities with active nightlife 
and subsequent day sports performance, we need to control for salient
features such as the ability of players on either team in the match-up,
whether any players are injured, the momentum of success each team is carrying
in the season, etc. To that end, we utilize spread data.\citep{anderson} The benefit of using spread data is that it controls for a lot, but the downside is
we don't know exactly what it controls for.

\begin{figure}[ht]
  \centering
  \includegraphics[scale=0.65]{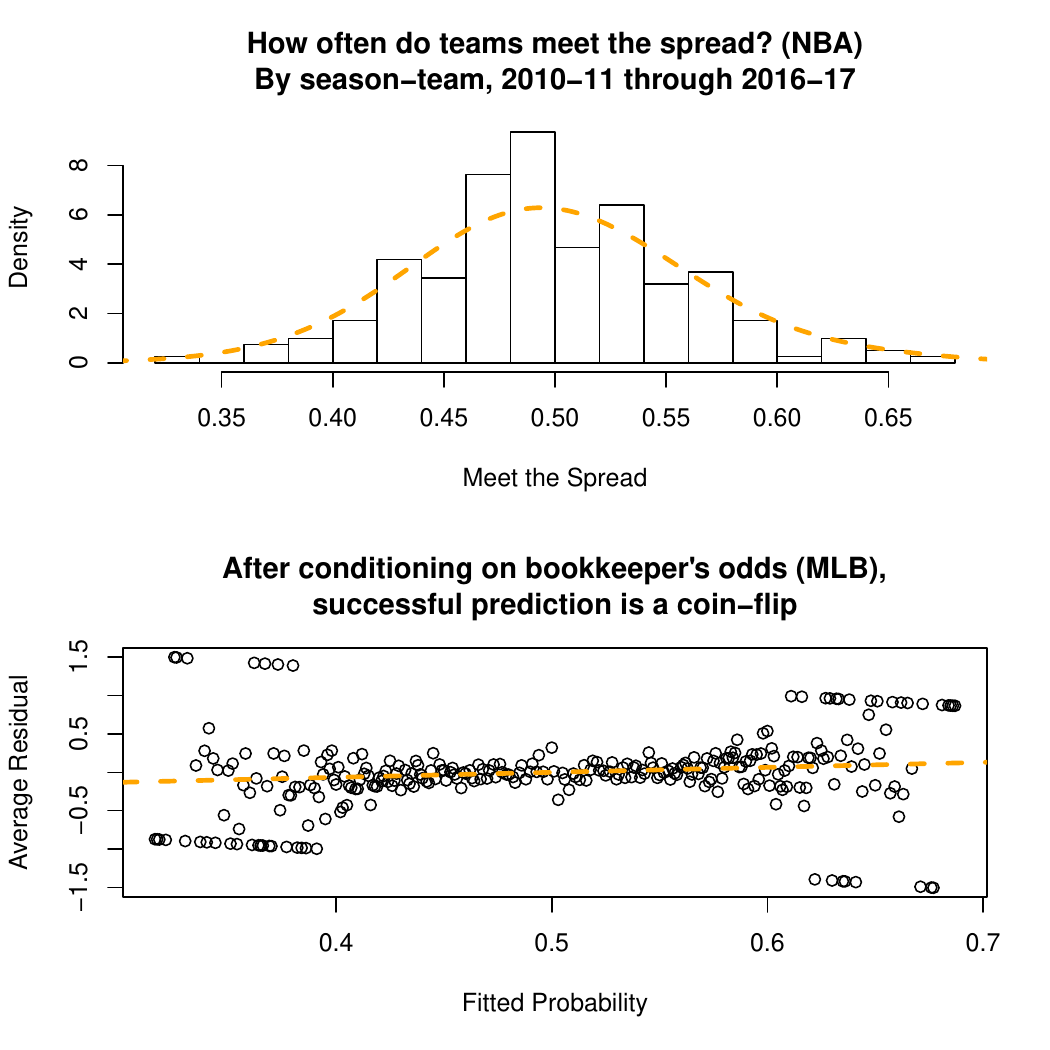}
  \label{fig: meet the spread}
  \caption{Empirically, in NBA any given team is apt to meet the spread
half the time across a particular season; in MLB, after conditioning on bookmakers' odds of
success, successful prediction of game outcome is like a coin-flip. 
In general, systematic deviations represent
unexplained mechanisms. }
\end{figure}

\paragraph{How an unknown spread variable may (not) affect our inference}
For example, if bookkeepers
are already aware of the effects of night-life on performance, it could
already be factored into the spread; if this were the case, then our variable
of interest would not appear to have significant explanatory power
after conditioning on spread, in spite of there being
a true effect. Alternatively,
if the house bookkeeper systematically mis-values
teams which are more likely to be encountered after playing a city with active nightlife,
we could see a spurious effect. 
That said, the amount of money that is at stake and the effort put into creating these spreads 
suggests that they are reasonably reliable. 
So, if we do find an effect in our variable of interest,
we can be reasonably confident in our conclusion that there is a causal effect and not some other systematic bias.

\subsection{BLS data} We also obtained data from
\href{https://www.bls.gov/data/}{Bureau of Labor Statistics} which records the number of establishments by business type at the MSA-quarter level. We obtain this data for each metropolitan statistical area corresponding
to an NBA or MLB team from years 2010-2016. 
As a proxy for how much night-life there is in a city, we look toward the 
number of sound recording studios, musical groups, and music publishers there 
are for a particular MSA and quarter: we simply calculate the total number of establishments across all three music categories listed above. We caveat that since the Toronto Raptors and Toronto Blue Jays are located in Canada, 
we don't have BLS data for these teams.\footnote{We don't expect Toronto, Canada to be a party city.}

We then merge this in with our lines data, taking care to do so according to each team's last game location such that the feature captures the proper
party effect. In addition, we lag our BLS data by one year for two reasons:
realistically, player expectations of which cities have the most active night-life may not
update in real-time and also doing so allows us to create a betting model, i.e.
we don't rely on any features we can't obtain before making a game-time prediction. 

\subsection{Minutes data for NBA}
We also gathered minutes data from \href{http://www.espn.com/nba/scoreboard/_/date/}{ESPN}.
The minutes data contains information on the number of free throws, field goals, and three pointers made and attempted by each player in each game.
For each season and each team, we have nearly complete 
minutes data for all games played: 85\% of our season-teams' have no more than 3 missing games.

\subsection{Location data} We are also interested
to account for the effects of jet-lag. For each team, we queried
a maps search engine to obtain a longitude and latitude describing their
stadium location.\citep{ggmap}

\begin{figure}
  \centering
  \includegraphics[scale=0.425]{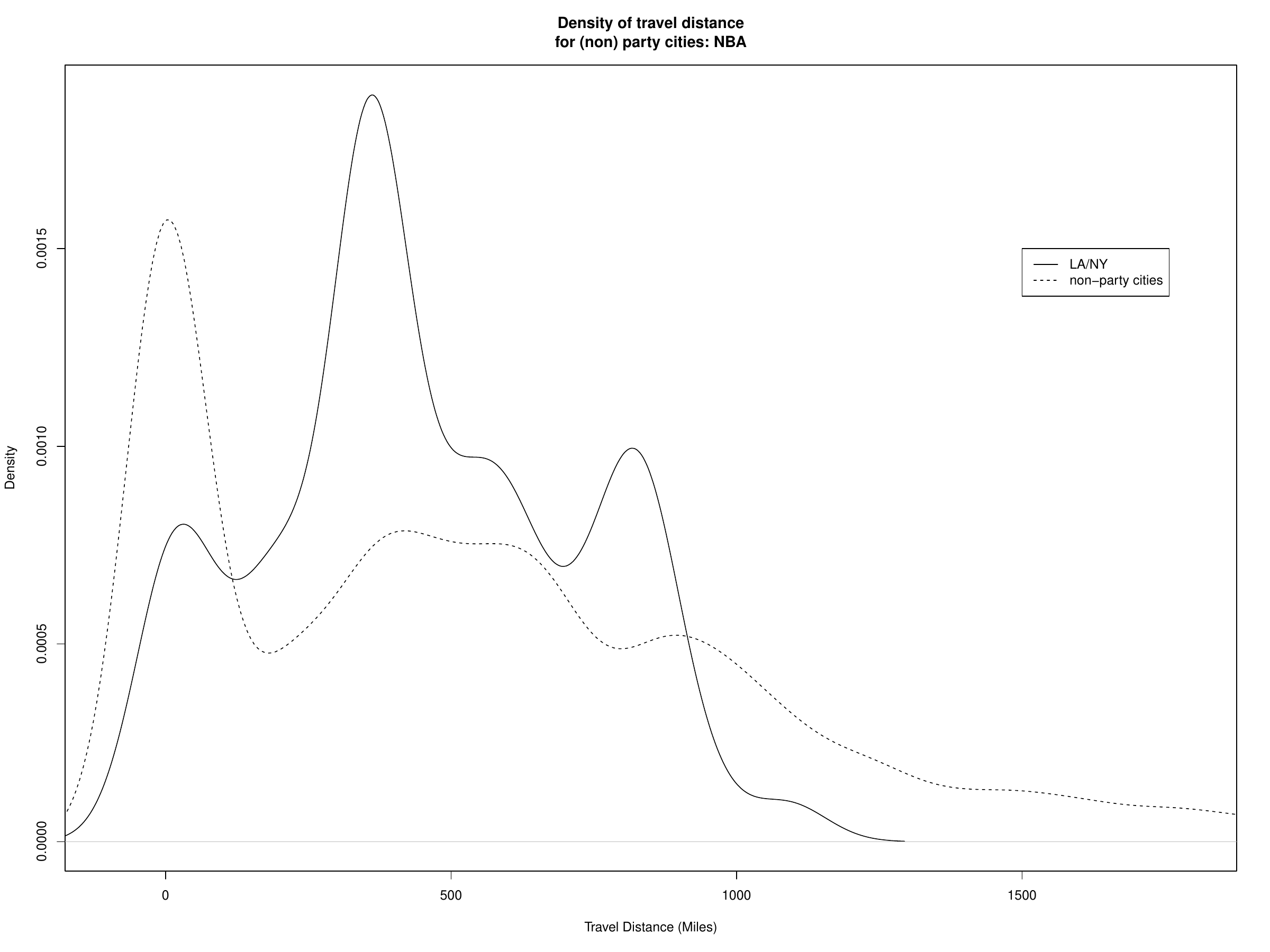}
  \label{fig: density by party}
  \caption{Within the NBA, we examine travel distance to current game location as a function of
    whether the last game was in a (non) party city. We see that in general,
    when traveling from a non party-oriented city to next game location, the distribution
    of travel-distance has fatter tails. This makes sense: as we saw in our previous
    heat-map, next day opponent geographically correlated with current game location.
    Since LA and NY are on the coast, their next day opponent is likely to be nearby; compare
    this to inland cities, for example, who may have to travel farther to reach their 
    nearest neighbor. We mention that the same     plot is not interesting to look at within the MLB, since games are played in     series, and therefore most travel-distances between games are zero miles.}
\end{figure}

\begin{figure}
  \centering
  \label{fig: party by team loc}
  \includegraphics[scale=0.6,angle=270]{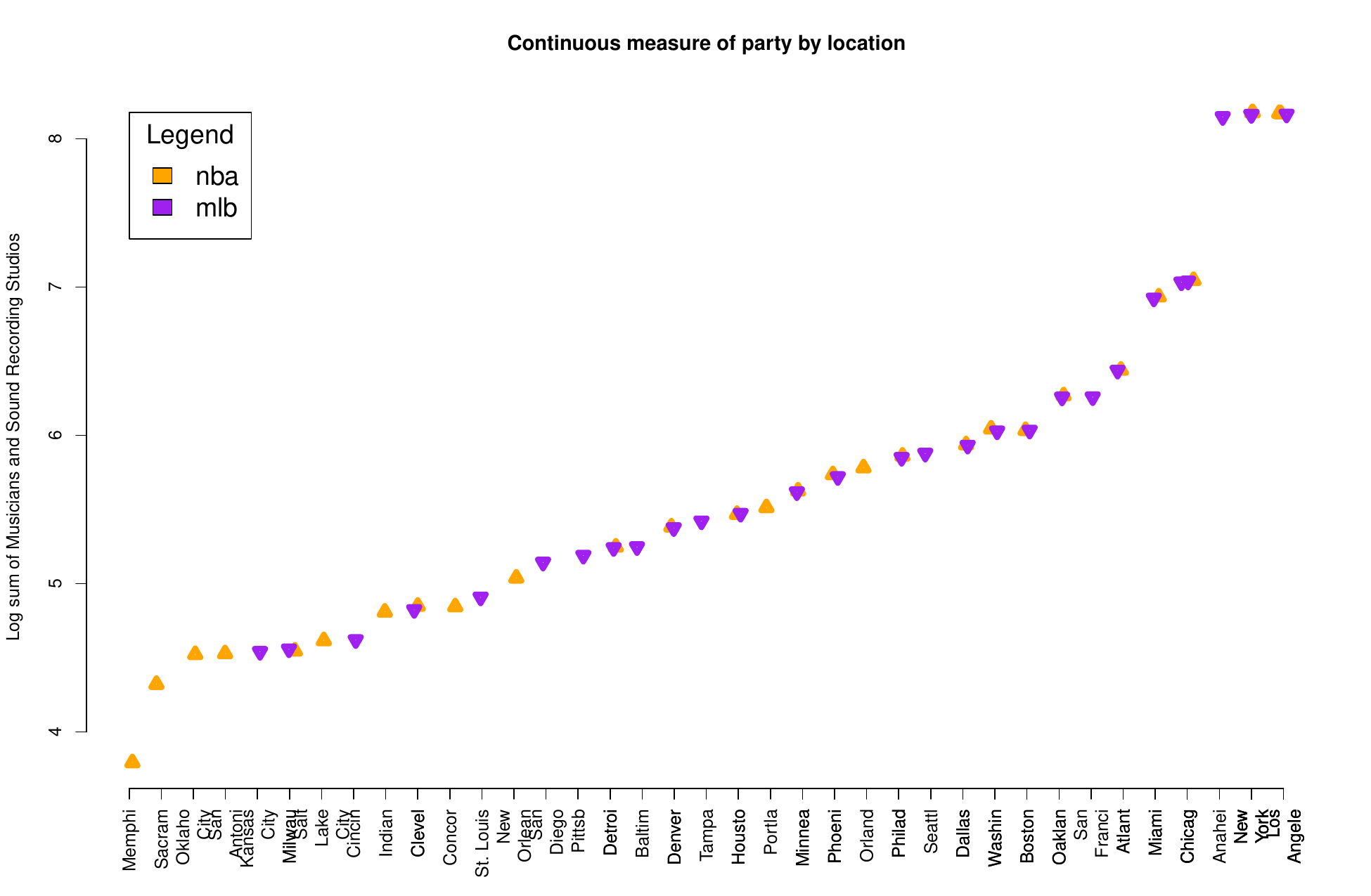}
  \caption{We plot our continuous measure of nightlife for each city within the NBA and MLB. We add a small amount of random noise to our x-coordinate such that cities
with both an NBA and MLB team are visible.}
\end{figure}

\section{Analysis of Team Performance}

\subsection{Modeling approach}

\paragraph{Home team effect}
Home-team advantage is well-studied.\citep{jones07, jones08}  
Empirically, teams play about half their games on the road, and half of their games at home. However,
it's possible that when a team plays on the road, their next game is more likely to be
an away game as well. I.e. most games after playing in a party city will be an away game.
This should be accounted for in the points-spread, but we also include
a home-team indicator to be sure.

\paragraph{Fatigue from previous game}
What if we pick up effects of a busy play schedule?
Aside from accounting for how much rest time a team is allowed in between games,
we're also interested to get a measure of
fatigue incurred on the court.
We account for this by bringing in minutes data, and counting the number of 
times the players run back and forth on the court. This is crudely measured
by the number of changes in posession. We back out this quantity by looking
at the number of three pointers and field goals made, the number of missed shots,
as well as the number of turnovers and rebounds.

\paragraph{Fatigue from jetlag}
What about jetlag? Prolonged air travel has in general been shown
to have a negative effect on sports performance.\citep{leeandgalvez}
Previous studies have shown that jetlag is a strong enough effect
to erase home field advantage in major league baseball.\citep{songetal}
To account for this, we calculate the travel distance between stadiums,
consider a measure for east-west travel (which induces a change in time-zone).

Specifically, for each pairwise set of coordinates, we calculated distance between the two
points.\citep{sp} To account for jet-lag, we need to not only consider travel distance
but the \emph{direction} of travel: if it's west to east the players' must adjust their internal
clock forward and hence they lose sleep, if it's east to west then the players gain rest time.
Hence we are interested to interact some continuous measure of east-west travel with
travel-distance in order to come up with a measure for jet-lag.
To that end, we first computed the final direction 
(bearing) when arriving at the 
second point after starting from the first and following the shortest path on a sphere 
(following a great circle), which yields a direction in degrees. 
We then obtain a measure for east-west travel by looking at the trigonometric-$\sin$ of this angle after converting to radians.
Interacting this with travel-distance gets us a measure of jetlag.\citep{lallensack}

\paragraph{Model} We use a logistic regression to estimate the likelihood of meeting
the spread as a function of whether the team partied as well as controlling for other confounders
listed above. Our basic model accounts for the rest-time since
last game, jet-lag, home-team effect, player fatigue incurred during the last game, and the party effect.

\subsection{Results}
\subsubsection{NBA: using points-spread data}
We find that after accounting for player fatigue incurred during the last game, jetlag and travel fatigue, recovery time, and home-team effect that both our discrete indicator and continuous measure of party
have a statistically significant effect on likelihood of meeting the spread.
In laymens terms, teams under perform bookmaker's expectations after playing in cities 
with active nightlife.

\begin{tabular}{@{\extracolsep{5pt}}lcc}  \\[-1.8ex]\hline  \hline \\[-1.8ex]   & \multicolumn{2}{c}{\textit{Dependent variable:}} \\  \cline{2-3}  \\[-1.8ex] & \multicolumn{2}{c}{Meet the Spread} \\  \\[-1.8ex] & (1) & (2)\\  \hline \\[-1.8ex]   Party discrete & $-$0.557$^{***}$ &  \\    & (0.149) &  \\    & & \\   Party continuous &  & $-$0.158$^{*}$ \\    &  & (0.095) \\    & & \\   Lag changes in posession & 0.189 & 0.181 \\    & (0.140) & (0.140) \\    & & \\   Logged travel distance & $-$0.015 & $-$0.010 \\    & (0.026) & (0.026) \\    & & \\   East-west travel direction & $-$0.072 & $-$0.037 \\    & (0.234) & (0.234) \\    & & \\   Number hours rest time & 0.001 & 0.001 \\    & (0.001) & (0.001) \\    & & \\   Time of game during day & 0.006 & 0.004 \\    & (0.013) & (0.013) \\    & & \\   Home team effect & 0.002 & 0.004 \\    & (0.044) & (0.044) \\    & & \\   Logged travel distance * east-west & 0.011 & 0.007 \\    & (0.036) & (0.036) \\    & & \\   Constant & $-$0.923 & $-$0.875 \\    & (0.695) & (0.695) \\    & & \\  \hline \\[-1.8ex]  Observations & 9,517 & 9,517 \\  Log Likelihood & $-$6,586.320 & $-$6,592.159 \\  Akaike Inf. Crit. & 13,200.640 & 13,212.320 \\  \hline  \hline \\[-1.8ex]  \textit{Note:}  & \multicolumn{2}{r}{$^{*}$p$<$0.1; $^{**}$p$<$0.05; $^{***}$p$<$0.01} \\  \end{tabular}  

This model supports the hypothesis that players enjoy nightlife systematically in certain
cities and that a ``hangover'' effect is felt on the court the next day.
The features describing jet-lag and recovery time are not significant, but this is to be
expected: such salient information has already been taken into account in setting the spread.
In the next section, we explore which specific areas of gameplay are affected.

\subsubsection{MLB: using money-lines data}
We replicate the same analysis with
Major League Baseball, which is also well set up for our quasi-experiment since
there are many games played back-to-back in the season (more in fact, than NBA).
There are also anecdotes of players celebrating with alcohol in public.\citep{miller,campbell}

\paragraph{Differences between NBA and MLB as it pertains to our analysis}
In the MLB, money-lines
are used instead of point-spreads. So as opposed to predicting
the likelihood of meeting the spread, we predict the probability
a team would win as both a function of the bookmakers'
odds of winning and our party variable. 
Additionally, our methodology changes ever so slightly: when creating our
continuous measure of nightlife, we interact the variable with weekend. Interviews with athletes suggest that MLB players are more likely to
restrict their partying to weekends.
One of the last major differences between NBA and MLB games is that MLB games
are played in series, so the same match-up is repeated multiple times. This remedies
the problem of an unknown travel schedule we faced with NBA because we are sure 
teams stay overnight in the same location within a series.

\begin{tabular}{@{\extracolsep{5pt}}lcc}  \\[-1.8ex]\hline  \hline \\[-1.8ex]   & \multicolumn{2}{c}{\textit{Dependent variable:}} \\  \cline{2-3}  \\[-1.8ex] & \multicolumn{2}{c}{Probability of Winning} \\  \\[-1.8ex] & (1) & (2)\\  \hline \\[-1.8ex]   Continuous measure of nightlife & $-$0.120$^{*}$ &  \\    & (0.071) &  \\    & & \\   Nightlife (no weekend interaction) &  & 0.137 \\    &  & (0.134) \\    & & \\   Bookmaker's odds & 2.682$^{***}$ & 2.669$^{***}$ \\    & (0.158) & (0.158) \\    & & \\   Home-team effect & 0.014 & 0.144 \\    & (0.034) & (0.102) \\    & & \\   Number of rest days & 0.018 & 0.016 \\    & (0.021) & (0.021) \\    & & \\   Logged travel distance & $-$0.005 & $-$0.003 \\    & (0.005) & (0.005) \\    & & \\   Weekend & 0.044 & 0.002 \\    & (0.038) & (0.029) \\    & & \\   Constant & $-$1.363$^{***}$ & $-$1.468$^{***}$ \\    & (0.084) & (0.122) \\    & & \\  \hline \\[-1.8ex]  Observations & 26,473 & 26,473 \\  Log Likelihood & $-$18,124.350 & $-$18,125.300 \\  Akaike Inf. Crit. & 36,272.700 & 36,264.600 \\  \hline  \hline \\[-1.8ex]  \textit{Note:}  & \multicolumn{2}{r}{$^{*}$p$<$0.1; $^{**}$p$<$0.05; $^{***}$p$<$0.01} \\  \end{tabular}  

As shown above, our continuous measure of nightlife has a statistically significant negative effect for MLB teams likelihood of winning.

\subsubsection{Betting performance}
We also validate our MLB model by seeing how well we can perform against the
market.
\begin{figure}[!h]
  \centering
  \label{mlb bets by season}
  \includegraphics[scale=0.82]{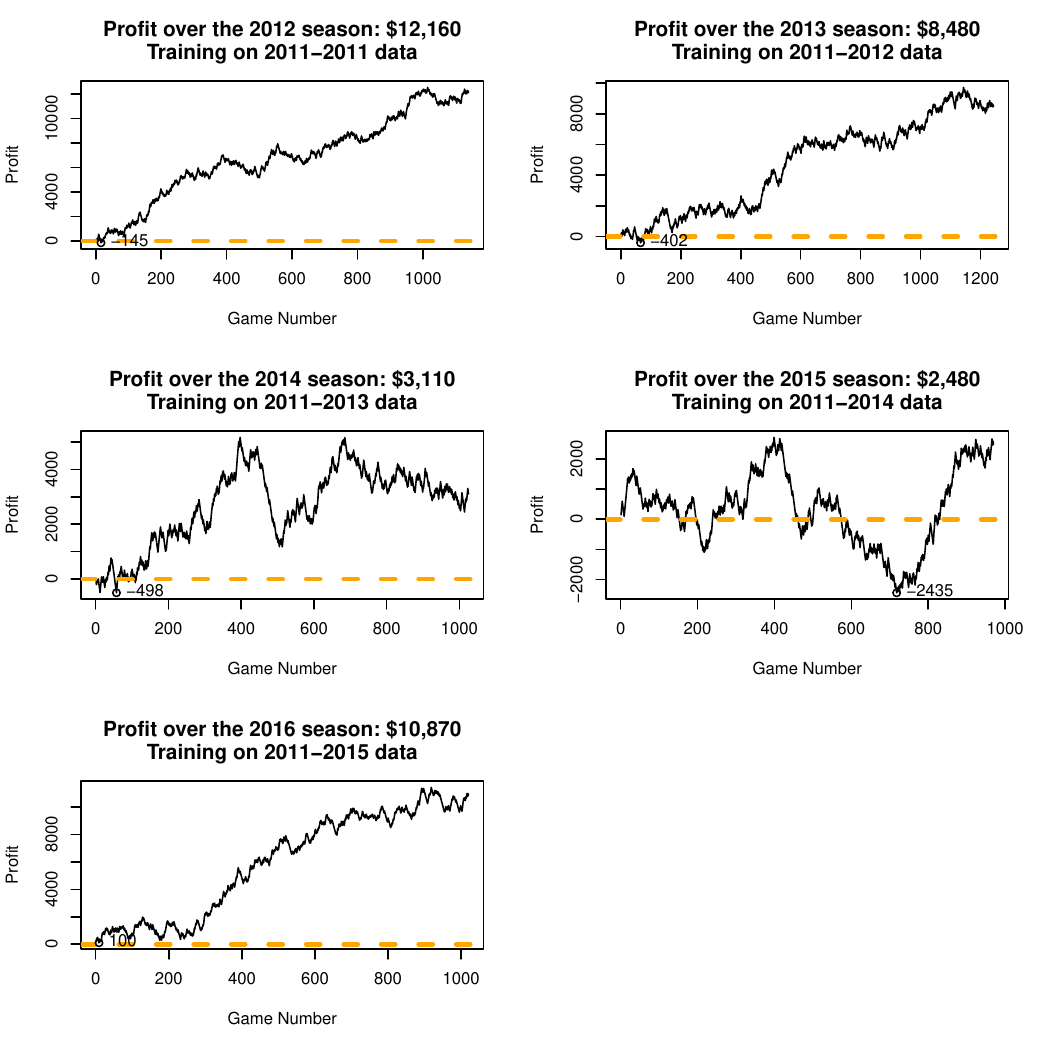}
  \caption{Betting performance by season. In each season, the 
    circled point represents the worst out of pocket expense incurred.
    We first train a model on historical data then use said model to predict on ``current year''
    data, plotting profits as a function of time over the ``current'' season. For each relevant
    game, we place a bet of \$100 if the expected value of the bet is positive after accounting for the bookmaker's cut. E.g. in the early 2016 season,
    at its worst we are out of pocket \$89 dollars. By season end, we have     recouped all of
    our initial investment plus an additional \$11.5k. We are profitable in each season, and mention that 2015 is the ``worst'', but we still come out ahead.}
  \label{bettingperf}
\end{figure}

\subsubsection{Placebo tests} As a sanity check with our NBA data, we interact a
discrete-indicator for Los Angeles and New York with an 
indicator for \emph{more} than 24 hours of rest
between games, and call this our placebo effect. The only difference between this feature
and our discrete indicator for nightlife is the time-window of interest.
The idea is that with sufficient rest time, the hangover effect dissipates.

Indeed, we see that if players have enough recovery time after visiting cities such as Los
Angeles or New York, we cannot reject the null hypothesis that player performance is unaffected
by visitation to cities with active nightlife. The analagous placebo test
for MLB is not relevant, since almost all games are played back to back.

\begin{tabular}{@{\extracolsep{5pt}}lc}  \\[-1.8ex]\hline  \hline \\[-1.8ex]   & \multicolumn{1}{c}{\textit{Dependent variable:}} \\  \cline{2-2}  \\[-1.8ex] & Meet the Spread (NBA) \\  \hline \\[-1.8ex]   Party placebo & 0.053 \\    & (0.084) \\    & \\   Lag changes in posession & 0.170 \\    & (0.140) \\    & \\   Logged travel distance & $-$0.004 \\    & (0.026) \\    & \\   East-west travel direction & $-$0.034 \\    & (0.235) \\    & \\   Number hours rest time & 0.001$^{*}$ \\    & (0.001) \\    & \\   Time of game during day & 0.002 \\    & (0.013) \\    & \\   Home team effect & $-$0.006 \\    & (0.044) \\    & \\   Logged travel distance * east-west & 0.006 \\    & (0.036) \\    & \\   Constant & $-$0.865 \\    & (0.695) \\    & \\  \hline \\[-1.8ex]  Observations & 9,517 \\  Log Likelihood & $-$6,593.344 \\  Akaike Inf. Crit. & 13,214.690 \\  \hline  \hline \\[-1.8ex]  \textit{Note:}  & \multicolumn{1}{r}{$^{*}$p$<$0.1; $^{**}$p$<$0.05; $^{***}$p$<$0.01} \\  \end{tabular}  
\subsection{Possible confounders or potential issues}

\paragraph{The problem of an unknown travel schedule} In NBA, we have the difficulty of not observing when teams travel. E.g. if a particular team  plays
in Los Angeles on Friday and another game in Boston the following day, we don't
know whether the team traveled on Friday evening or Saturday morning and hence don't know
which city they ``visited'' on Friday night. Heuristics for
travel schedules can be made, but it is hard to obtain reliable team-travel data.
In MLB, on the other hand, each match-up is actually a series of games. E.g.
if the Dodgers play the Giants in San Francisco, they will play several games spread
across several days, each in the same city; in these cases, we can infer that the team
is in fact spending the night in said city.

\section{Analysis of Player Performance}
Within the NBA, we additionally looked at actual performance metrics instead of relying on point-spread data. 
This allows us to avoid using a black-box feature, and also pin-point
more precisely what elements of the game are affected by visiting a nightlife city. This motivates looking at number of points allowed, and number of points scored; we also considered how rebounds, fouls, and even injuries might be 
explained (in part) by a hangover effect.

\paragraph{The hangover correlates with allowing more points}
Common knowledge suggests that the overall defense of a team
is effected strongly by individual player effort. It seems likely that a hangover effect
would be more prevalent on the defensive aspects of gameplay. To that end, we examine points allowed and points scored as a function of exposure to nightlife.
It's worth noting that travel to party cities is correlated with allowing
more points on the subsequent day of gameplay, but the null hypothesis
can't be rejected for points scored.

It's also interesting to see that after visiting LA or NY, for any given
team their next day performance is strongly correlated with allowing the other team to score two and a half points more. 
It's even more interesting to note that points spreads are on a similar order of magnitude, i.e. this could often be the difference between meeting the spread or not.

\hspace{-35pt}
\begin{tabular}{@{\extracolsep{5pt}}lccc}  \\[-1.8ex]\hline  \hline \\[-1.8ex]   & \multicolumn{3}{c}{\textit{Dependent variable:}} \\  \cline{2-4}  \\[-1.8ex] & \multicolumn{2}{c}{Team Points Admitted} & Team Points Scored \\  \\[-1.8ex] & (1) & (2) & (3)\\  \hline \\[-1.8ex]   Discrete party indicator & 2.970$^{***}$ &  &  \\    & (0.835) &  &  \\    & & & \\   Continuous party measure &  & 1.661$^{***}$ & $-$0.139 \\    &  & (0.644) & (0.637) \\    & & & \\   Lag change in possessions & 6.694$^{***}$ & 6.751$^{***}$ & 5.858$^{***}$ \\    & (0.973) & (0.973) & (0.963) \\    & & & \\   Logged travel distance & $-$0.114 & $-$0.132 & $-$0.285 \\    & (0.198) & (0.198) & (0.196) \\    & & & \\   Number hours since last game & 0.011$^{*}$ & 0.015$^{**}$ & 0.016$^{**}$ \\    & (0.006) & (0.007) & (0.007) \\    & & & \\   Home team effect & $-$3.269$^{***}$ & $-$3.269$^{***}$ & 2.417$^{***}$ \\    & (0.303) & (0.303) & (0.300) \\    & & & \\   Constant & 68.114$^{***}$ & 67.686$^{***}$ & 71.348$^{***}$ \\    & (4.805) & (4.810) & (4.758) \\    & & & \\  \hline \\[-1.8ex]  Observations & 6,234 & 6,234 & 6,234 \\  R$^{2}$ & 0.091 & 0.090 & 0.084 \\  Adjusted R$^{2}$ & 0.087 & 0.086 & 0.079 \\  Residual Std. Error (df = 6201) & 11.670 & 11.675 & 11.549 \\  F Statistic (df = 32; 6201) & 19.453$^{***}$ & 19.247$^{***}$ & 17.720$^{***}$ \\  \hline  \hline \\[-1.8ex]  \textit{Note:}  & \multicolumn{3}{r}{$^{*}$p$<$0.1; $^{**}$p$<$0.05; $^{***}$p$<$0.01} \\  \end{tabular}  
\section{Conclusions and Avenues for Future Research}
\subsection{Extending the research}
\paragraph{Within the NBA} 
We're interested to explore what other factors may affect a team's propensity to suffer hangover effects beyond simply visiting a party city; e.g. are teams more likely to go out after
a win or a loss? Do teams who aren't going to make the playoffs party more? Are there demographic
factors which influence partying, e.g. are younger, unmarried players more susceptible to treatment effect? Within a game, it would be interesting to look at 
number of shots taken, and number of shots taken by 
stars as a function of previous day visitation to party cities. We could also investigate 
effects on injury risk and flagrant foul rates.

\paragraph{MLB}
It would also be interesting to examine the particular ways that partying
effects performance statistics similar to how we did with NBA.
We might consider whether batting performance (averages, slugging percentage)
pitching (ERA, WHIP) or general fielding stats are effected by party.

\paragraph{General}
In both sports, it's possible there are prolonged effects of partying, and applying some sort of kernel on the number of hours played since the last game could be interesting; recall that our decision to only look at back-to-back games 
within 24 hours of each other was only motivated by intuition and not guided by data; How does the effect taper off as a function of time? 
Future work may consider similar effects in NHL or European soccer leagues.

\subsection{Conclusions}
We do find that visiting a city with active nightlife the day before a game does have 
a statistically significant negative impact on a team's likelihood of meeting the spread in NBA. This decreased likelihood to meet the spread may be explained 
by the team's defense letting up more points. In turn, we believe that the drop in
defensive performance may be caused by a drop in physical exertion.
We verify that this phenomena persists across sport, and 
to that end look toward major league baseball money-lines. Using these, we back out gambling
house's estimate for the team's probability of winning the game, and after conditioning on this
we find that visiting a city with active nightlife the day before a game does negatively 
impact the team's likelihood of winning.
With a causal model established, we realize a profitable betting strategy in each year of testing.

\bibliographystyle{DeGruyter}
\bibliography{sample}

\section{Appendix}
\begin{figure}[h]
  \centering
  \hspace{-55pt}
  \label{fig: next day opponent mlb}
  \includegraphics[scale=0.45]{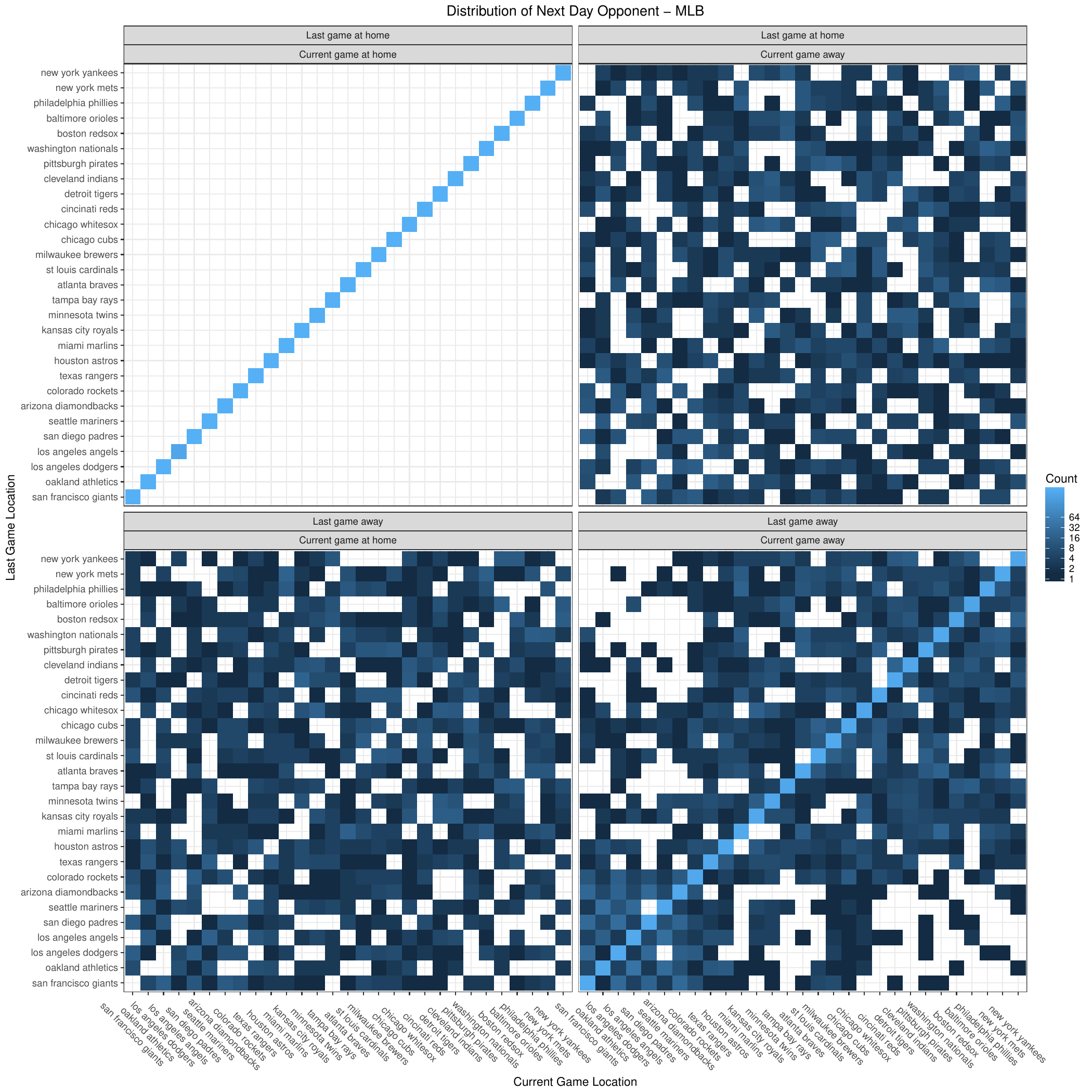} 
  \caption{For the MLB, there are some differences when compared with the     analogous
  plot for NBA. In particular, notice that a common scenario is for a team to play two away games 
  back to back at the same location. This is an artifact of baseball match-ups being played 
  in series as opposed to one-off games like in the NBA. We also notice in back-to-back
  away games that the sparse rows 
  (i.e. locations for which next-day opponent is restricted to only a strict subset of the 
  29 available teams to play) all are attributed by West coast teams. In particular, 
  Washington Nationals, San Francisco Giants, Oakland Athletics, Los Angeles Dodgers, and 
  Arizona Diamondbacks are locations from which a next-day game on the East coast are not possible.}
\end{figure}

\begin{sidewaysfigure}   
  \centering
  \label{fig: why drinks or pop dont work}
  \includegraphics[scale=0.5]{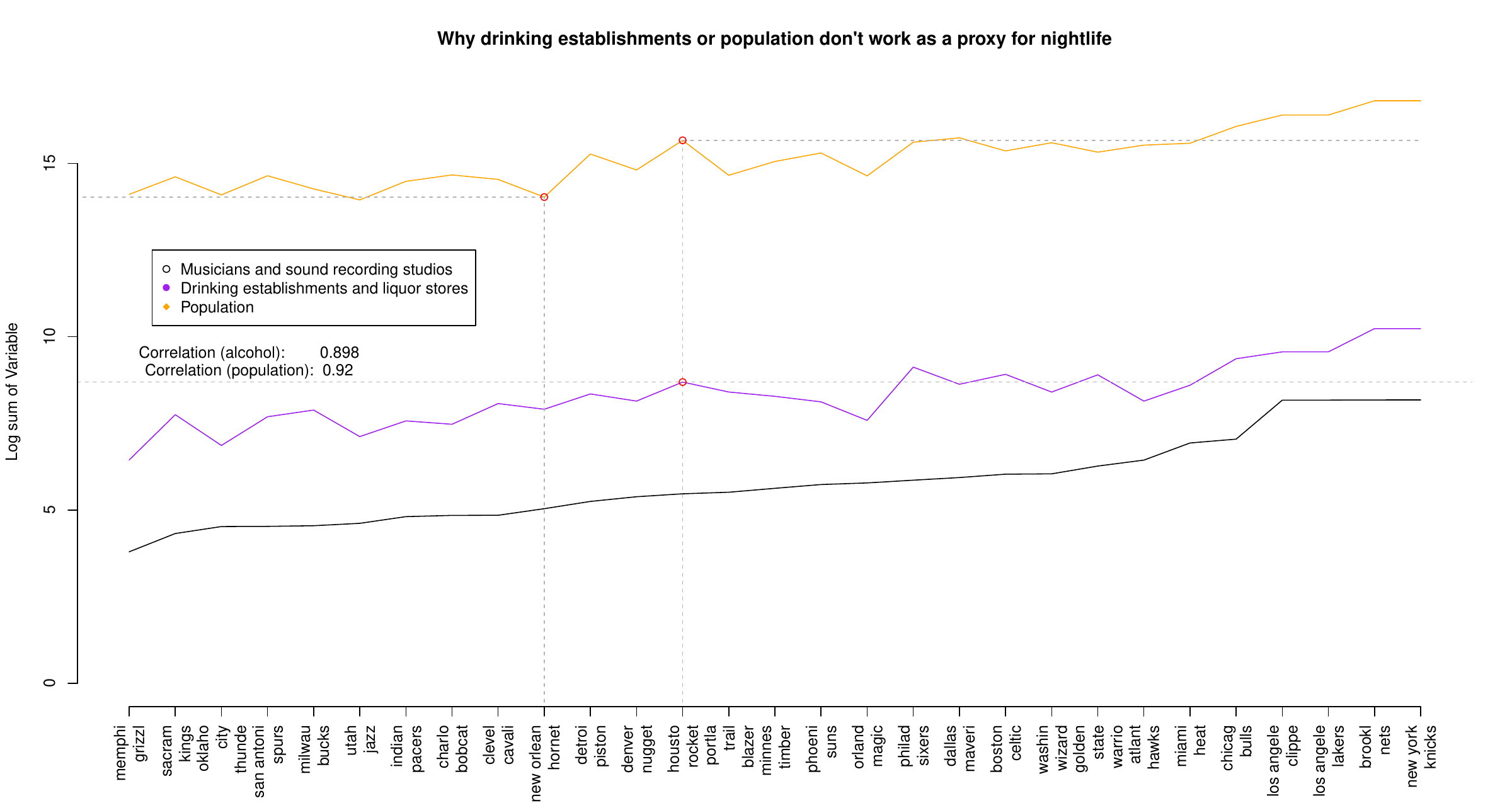} 
  \caption{We remark why using a statistic such as number of drinking establishments or 
    population size doesn't work. Although both of these are linearly correlated with musicians, 
    and even in spite of the extremal observations being ranked consistently across each statistic, 
    i.e. Memphis toward the bottom and Los Angeles more or less at the top,  
    the problem is that population and number of drinking establishments have sinusoidal 
    relationships with respect to the rank ordering imposed by the musicians index 
    that do not properly reflect variation in nightlife between cities. For example, both 
    incorrectly rank San Antonio, Texas as having one of the higher nightlife indices. Another example, 
    population incorrectly places New Orleans, Louisiana as having a low     nightlife index, when in reality it is famous for its vibrant city life.}
\end{sidewaysfigure}

\end{document}